\documentstyle{article}
\def\beq{\begin{equation}}
\def\eeq{\end{equation}}

\begin{document}
\large
\title{THE STATIC GRAVITATIONAL FIELD OF A SPHERICALLY SYMMETRIC BODY}
\author{M. Novello - \\
Centro Brasileiro de Pesquisas Fisicas \\
Luciane R. de Freitas \\
Instituto de Fisica, UFRJ \\
and \\
V. A. De Lorenci \\
CBPF \\
Rio de Janeiro}
\date{ }
\maketitle

\begin{abstract}
We continue here the exam \cite{LuMa} of a theory of gravity that satisfies
the Einstein Equivalence Principle (EEP) for any kind of matter/energy, except
for the gravitational energy. This is part of a
research program that
intends to re-examine the standard Feynman-Deser approach of field
theoretical derivation of Einstein\rq s General Relativity. The hypothesis
implicit in such precedent derivations \cite{Feynman} \cite{Deser}
concerns the universality of gravity
interaction. Although there is a strong observational basis supporting the
universality of matter to gravity interaction, there is not an
equivalent situation that supports the hypothesis that gravity interacts
with gravity as any
other form of non-gravitational energy. We analyse here a kind of
gravity-gravity interaction distinct from GR but, as we shall see, that
conforms with the actual status of observation. We exhibit the
gravitational field produced by a spherically symmetric static configuration
as described in this field theory of gravity. The values that we obtain for
the standard PPN parameters ($\alpha = \beta = \gamma = 1$) coincide
with those of General Relativity. Thus, as we pointed out in
a previous paper, the main different aspect of our theory and GR concerns the
velocity of the gravitational waves. Since there is a large expectation that
the detection of gravitational
waves will occur in the near future, the question of which theory describes
Nature better will probably be settled soon.
\end{abstract}

\section{INTRODUCTION}
\subsection{A. Introductory Remarks}

In a recent paper two of us (LRF and MN) have proposed a critical
re-examination of the general treatment concerning
Einstein\rq s Equivalence Principle (EEP). The starting
point rests on the well-known fact that EEP is
supported by experiment by at least\footnote{See the status of the
STEP (Satellite Test of the Equivalence Principle), an international
tentative of improving this result to one part in $10^{17}$.} one part in
$10^{11}$. This
statement should be understood in the very limited sense that it concerns
observed effects of gravity restrained to
the coupling of matter (that is, any non-gravitational form of energy) to the
gravitational field. However, many theories of gravity, starting from General
Relativity, go far beyond such limited domain of observation and
assume a generalized form which states that the EEP comprises absolutelly
all forms of energy (including gravitational).
This, of course, is a speculation that still nowadays remains
beyond any experimental test.
The recent general excitation among the scientific community concerning the
possible detection of gravitational waves led to the belief that the
observation of these waves could act as a crucial test on gravity-gravity
interaction and to provide for a decision concerning which,
among the rivals gravity theories, is the best one.

In \cite{LuMa} a class of
alternative models for gravity was presented and a specific one was
developed. Our purpose in the present paper is to proceed with such an
investigation on the consequences of our theory.
Here we will describe a solution of our set of equations of the gravitational
field produced by a spherically symmetric and static configuration. We shall
show that as far as the post-Newtonian approximation is concerned, the
behavior of matter (i.e. any form of non-gravitational energy), in
our theory is indistinguishable from General Relativity (GR).

{}From the experimental standard tests we can state the following
results\footnote{We are following the standard notation for the PPN
parameters. The reader not aware of this formulation should consult the
Appendix and the references.}

\begin{itemize}
 \item{$\alpha = 1$}
 \item{$\gamma = 1.000 \pm  0.002$}
 \item{$\frac{1}{3} (2 \gamma + 2 - \beta) = 1.00 \pm 0.02$}
 \item{$\chi < 10^{-3}$}
 \item{$ \alpha_{2}  < 4 x 10^{-4} $}
 \item{$\frac{2}{3} \alpha_{2} - \alpha_{1}  < 0.02$}
 \item{$49 \alpha_{1} - \alpha_{2} - 2.2 \chi  < 0.1$}
\end{itemize}

We shall prove in this paper that, for the new theory of gravity that we
are examining, like in the very same manner as General
Relativity, the unique non vanishing parameters are
$\alpha$, $\beta$ and $\gamma$. The extra parameters, present in others
theories describe strange non-usual properties of the gravitational
interaction\footnote{Like, for instance, possible deviation from
Lorentz symmetry, effects of prefered frames, non conservation of
energy-momentum at the post-Newtonian limit and an eventual spatial
anisotropy of three-body interaction.}.

This result supports a remark made in our previous paper which states that
the crucial distinction of ours gravity theory and GR should be made
in evidence from the detection of the gravitational waves.

\subsection{B. Synopsis}
The plan of the paper is the following. In section \ref{field} we make a
very short review of the field theory of gravitation\footnote{The reader
that would like to know more about this should look for the references of the
quoted papers by Feynman, Deser and our previous paper.}. We present the
fundamental variables that describe the gravitational field and its
corresponding equation of motion. In section \ref{solution} we describe the
symmetries of the problem and find the correspondent equations of
motion. A solution for the static and spherically symmetric configuration
is then presented. In section \ref{effective} we show the
effective geometry that
is the observed effect of the previous solution as seen by any
non-gravitational form of energy. In section \ref{energy} we evaluate the
corresponding distribution of gravitational energy within our theory. We
end with section \ref{conclusion} in which some comments concerning
the present solution and its consequences are shown.

\section{Field Theory of Gravity: A Summary}
\protect\label{field}

In this section we will provide a short resum\'e of the theory of gravity
that we will analyse. Actually, we will deal here with a field
theoretical model of gravitational interaction. In the standard
traditional way this field is represented by a
symmetric second order tensor that we denote by $\varphi_{\mu\nu}$. This
field is supposed to propagate in an auxiliary background Minkowski
geometry\footnote{We shall see that this metric is not
observable neither by matter nor by the gravitational field.}.

In order to exhibit the general covariance of the theory we will
write the auxiliary metric $\gamma_{\mu\nu}$  in an arbitrary
system of coordinates. We define the corresponding covariant derivative by
the standard way:

\beq
V_{\mu\hspace{0.1cm};\hspace{0.1cm}\nu} =
V_{\mu\hspace{0.1cm},\hspace{0.1cm}\nu} -
\Delta^{\alpha}_{\mu\nu} V_{\alpha}
\label{T1}
\eeq

in which
\beq
\Delta^{\alpha}_{\mu\nu} =
\frac{1}2 \gamma^{\alpha\beta}\hspace{0.1cm}(
\gamma_{\beta\mu\hspace{0.1cm},\nu} +
 \gamma_{\beta\nu\hspace{0.1cm},\mu} - \gamma_{\mu\nu\hspace{0.1cm},\beta}).
\label{T2}
\eeq

The associated curvature tensor vanishes identically that is

\beq
R_{\alpha\beta\mu\nu}(\gamma_{\epsilon\lambda}) = 0.
\label{T3}
\eeq

The gravitational field is represented by a three-index tensor
$F_{\alpha\beta\mu}$ obtained from the potential $\varphi_{\mu\nu}$
by derivation:

\beq
F_{\alpha\beta\mu} = \varphi_{\mu[\alpha;\beta]} +
\varphi_{,[\alpha}\gamma_{\beta]\mu} +
\gamma_{\mu[\alpha}{{\varphi_{\beta]}}^{\lambda}}_{;\lambda}.
\label{T4}
\eeq
where we are using the anti-symmetrization symbol $[\hspace{0.9mm}]$ like

$$ [A, B] \equiv AB - BA.  $$

We use an analogous form to the symmetrization symbol $(\hspace{0.9mm})$

$$ (A, B) \equiv AB + BA.  $$

{}From the above definition it follows that the field $F_{\alpha\beta\mu}$
is anti-symmetric in the first pair of indices and obeys the cyclic identity,
that is

$$ F_{\alpha\mu\nu} + F_{\mu\alpha\nu} = 0 $$
and

$$ F_{\alpha\mu\nu} + F_{\mu\nu\alpha} + F_{\nu\alpha\mu} = 0.$$

The trace of the tensor $F_{\alpha\beta\mu}$ is given by

\beq
F_{\mu} \equiv F_{\mu\alpha\beta} \gamma^{\alpha\beta} =
4({{\varphi_{\mu}}^{\lambda}}_{;\lambda} - \varphi_{,\lambda}).
\label{T5}
\eeq

This allow us to re-write the expression of the field in the
form

\beq
F_{\alpha\beta\mu} = \varphi_{\mu[\alpha;\beta]} +
 \frac{1}{4} F_{[\alpha}
\gamma_{\beta]\mu}.
\label{T6}
\eeq

The equation of motion of the gravitational field will appear in
a more convenient form when written in terms of
an associated quantity $M_{\alpha\beta\mu}$ that has the same
symmetries as  $F_{\alpha\beta\mu}$ and is defined by

\beq
M_{\alpha\beta\mu} \equiv F_{\alpha\beta\mu} -
\frac{1}2 F_{\alpha} \gamma_{\beta\mu} + \frac{1}2
F_{\beta} \gamma_{\alpha\mu}.
\label{d7}
\eeq

See \cite{LuMa} for other properties of these quantities.

The symbol $\kappa$ represents Einstein\rq s constant, written in
terms of Newton\rq s constant $G_{N}$ and the velocity of light $c$
by the definition $$ \kappa = \frac{8\pi}{G_{N} c^{4}}. $$ We set $c
= G_{N} = 1.$

Any theory of gravity that yields the correct weak field limit must
reduces in this limit to the standard Fierz equation of motion. We
have shown, in our previous paper, that in order to achieve such
correct limit the theory, described in terms of the gravitational
field $F_{\alpha\beta\mu}$, must be such that its Lagrangian is
a functional of the quantity $Z$ defined in terms of the invariants $A$ and
$B$ under the form:

\beq
Z \equiv A - \frac{3}{4} B.
\protect\label{Z1}
\eeq
in which
$$ A \equiv F_{\alpha\mu\nu} F^{\alpha\mu\nu} $$
and
$$ B \equiv F_{\alpha} F^{\alpha}. $$

We have argued that a good candidate for this is a theory the
Lagrangian of which is given by

\beq
L = \frac{b^{2}}{k} \left\{  \sqrt{1 + \frac{1}{b^{2}} ( -A +
\frac{3}{4} B )} - 1  \right\}.
\protect\label{Z2}
\eeq
We will analyse here only this particular form among the whole class of
theories that could be constructed with $Z$.

The equation of motion that follows from this Lagrangian is then given by

\beq
{G^{L}}_{\mu\nu} = \frac{1}{2} {L_{A}}^{-1} \left\{ L_{A;\lambda}
{M^{\lambda}}_{(\mu\nu)}  +  {T^{m}}_{\mu\nu}   \right\}
\protect\label{Z3}
\eeq
in which we have added the source term represented by the
energy-momentum tensor of matter.

In the theory of General Relativity the corresponding equation of
motion of the gravitational field in the geometrical representation takes the
form:

$$ R_{\mu\nu} - \frac{1}{2} R g_{\mu\nu} = -  {T^{m}}_{\mu\nu} $$
in which the curvature of the associated Riemannian metric appears
explicitly. However there is an equivalent way to describe the
equation of motion in GR that seems worth to mention here just in order to
exhibit the similarity and the distinction between both theories.
Indeed, many authors (see, for instance \cite{Deser})  have shown that it is
possible to re-write Einstein\rq s equation of motion by
isolating the linear operator ${G^{L}}_{\mu\nu}$ from the non-linear
terms, showing explicitly the form of the self-interacting terms. It
is given by:
\beq
{G^{L}}_{\mu\nu} =  -  {T^{m}}_{\mu\nu} -  t_{\mu\nu}
\protect\label{Z4}
\eeq
in which  $t_{\mu\nu}$ is a complicated non-linear expression
constructed with the metric tensor and its derivative\footnote{The
reader not acquainted with this formulation should consult \cite{GPP}
for a very didactic presentation of it.}
A simple inspection on both theories thus exhibits very clearly the
particular characterization of the self-interacting terms in each
theory. Let us emphasize that in both cases of equations (\ref{Z3})
and (\ref{Z4}) they represent the full theory. There is no
approximation of any sort in these expressions.

Let us just make one more comment on the properties of these rivals theories
of gravity. Since both theories satisfy the restricted EEP, the
behavior of matter (or any form of non-gravitational energy) is precisely the
same in both theories. Matter follows geodesics in an effective Riemannian
geometry. Only the gravity-gravity processes are distinct. Since we are not
dealing in the present paper with gravitational waves we will not
consider any longer such distinction here.

So much for the review. Let us now look for a solution of our
equation (\ref{Z3}), in the absence of matter.

\section{THE SOLUTION}
\protect\label{solution}

We set for the auxiliary metric of the background the form

\beq
ds^{2} = dt^{2} - dr^{2} - r^{2} ( d\theta^{2} + sin^{2}\theta d\varphi^{2} )
\protect\label{s1}
\eeq
This means that all operations of raising and lowering indices are made by
this Minkowski metric $\gamma_{\mu \nu}$\footnote{The reader should be
attentive to the fact that matter (massive or massless particles --photons,
for instance -- that is, any form of non-gravitational energy) feels
a modified geometry. See the previous paper \cite{LuMa}.}.

We search for the simple solution in such a way that the only
non-identically null gravitational potential components
$\varphi^{\mu\nu}$ are only

$$ \varphi_{00} =  \varphi^{00} = \mu(r) $$
and
$$ \varphi_{11} =   \varphi^{11}  = \nu(r). $$

The trace  $\varphi$  is then given by
$$ \varphi = \mu - \nu.  $$

{}From this we obtain the gravitational field $F_{\alpha\beta\mu}$. The
only non-null terms are

$$ F_{100} = -2 ( \mu' + \frac{\nu}{r} )  $$

$$ F_{122} =  \mu' r^{2} + 3 \nu r   $$

$$ F_{133} = sin^{2}\theta F_{122} . $$
in which we have used a prime $'$ to symbolise the derivative with respect
to the radial variable $r$.

Thus the unique component of the trace that remains is $F_{1}$ which is
given by:

$$ F_{1} = - 4 ( \mu' + 2 \frac{\nu}{r} ) $$

{}From these we can evaluate the associated quantities

\begin{itemize}
 \item{The invariant $A$.}
 \item{The invariant $B$.}
 \item{The associated tensor $M_{\mu\nu\alpha}$.}
\end{itemize}

We obtain directly the following values:

\beq
A = - 4 \left\{ 3 {\mu^{'}}^{2} + 10 \frac{\nu \mu^{'}}{r} +
11 \frac{\nu^{2}}{r^{2}} \right\}
\protect\label{a1}
\eeq

\beq
B = - 16 \left\{ {\mu^{'}}^{2} + 4 \frac{\nu \mu^{'}}{r} +
4 \frac{\nu^{2}}{r^{2}} \right\}
\protect\label{a2}
\eeq

\beq
M_{100} = 2  \frac{\nu}{r}
\protect\label{a21}
\eeq
\beq
M_{122} =  -  r^{2} ( \mu^{'} +  \frac{\nu}{r} )
\protect\label{a22}
\eeq

\beq
M_{133} = sin^{2}\theta M_{122}.
\protect\label{a23}
\eeq

Under the above hypothesis of spherically symmetry of the solution,
there remain only two non-trivially satisfied equations, which are
given by

\beq
\left\{L_{A} \frac{\nu}{r} \right\}^{'}  +  2 L_{A} \frac{\nu}{r} = 0.
\protect\label{a3}
\eeq

and

\beq
\mu^{'} + \frac{\nu}{r} = 0.
\protect\label{a4}
\eeq

It seems worth to remark that the above set of equations (\ref{a3}) and
(\ref{a4}) are the same for any theory that satisfies the fundamental
condition, that is, for functionals only of the quantity $Z$ defined
by equation (\ref{Z1}).

Let us now especialize this for our theory characterized by
equation (\ref{Z2}).

We have
\beq
L_{A} = - \frac{b^{2}}{2}\hspace{0.9mm} ( b^{2} + L)^{-1}
\protect\label{a6}
\eeq

Substituting the above tentative form of a solution in the expressions
of the invariants we find for the quantity $Z$:
\beq
- A + \frac{3}{4} B = 4\hspace{0.9mm} \frac{\nu^{2}}{r^{2}}.
\protect\label{a7}
\eeq

Using this value in the equation (\ref {a3}) we find the solution

\beq
\nu = -\hspace{0.9mm} \frac{2M}{r}\hspace{0.9mm}
\left\{ 1 - {(\frac{r_{c}}{r}})^{4} \right\}^{-\frac{1}{2}}.
\protect\label{a8}
\eeq

in which the constant $r_{c}$ is given by

$$ {r_{c}}^{2}  \equiv \frac{4M}{b}. $$
The remaining function $\mu$ is given in terms of the elliptic
function $F(\alpha, \frac{\sqrt{2}}{2})$ by

\beq
\mu = \frac{1}{2} \sqrt{b M} \left\{ F(\alpha, \frac{\sqrt{2}}{2} ) +
\mu_{0} \right\}
\protect\label{a9}
\eeq
in which the constant $\mu_{0}$ must be chosen to yield the correct
assymptotic limit. The quantity $\alpha$ is given by

$$ \alpha \equiv arc sin
\left\{ \frac{cosh\hspace{0.5mm} x - 1}{cosh\hspace{0.5mm} x}
\right\}^{\frac{1}{2}} $$
and $x$ is defined by
$$ cosh\hspace{0.5mm} x \equiv \left\{ \frac{r}{r_{c}} \right\}^{2}. $$

At this point let us make a preliminary comment concerning the value of the
constant $b$. At the value $r = r_{c}$ the field has apparently a
singularity. We shall see in a subsequent section that we deal here
with a true singularity once the associated gravitational energy
is singular at point $r_{c}$. Although the quantity $b$ is a free
parameter of the theory, its value must be such that it provides that
the domain $r < r_{c}$ should be hidden inside
all known celestial bodies. A typical example associates $b$ to the inverse of
Planck length. This provides a good possible value that satisfies the above
requirement. We leave the actual value of $b$, for the time being, as an open
parameter, the true value of which is to be decided later on.

\subsection{The Effective Geometry}
\protect\label{effective}
In this section we will turn our attention to a geometric representation of
the present theory. Following the standard procedure \cite{Feynman}
\cite{Deser} we define a
Riemannian metric tensor in terms of the gravitational potential
as\footnote{The reader should note that one can make different forms of
geometric representation.
For instance one can uses the contravariant representation to set
$g^{\mu\nu} =  \gamma^{\mu\nu} + \varphi^{\mu\nu}$; or use pseudo
quantities by means of the determinant $\gamma$ in the definition of
the geometry. Each one of these choices provides non-equivalent
representations.}

\beq
g_{\mu\nu} \equiv \gamma_{\mu\nu} + \varphi_{\mu\nu}
\protect\label{b1}
\eeq
It seems worth to call attention to the reader that this definition has
a deep meaning, once for all forms of non-gravitational energy the net
effect of the gravitational field is felt precisely as if gravity
was responsible of changing the metrical properties of the spacetime from the
flatness structure to a curved one that is related to the gravitational field
precisely by the above expression\footnote{The
reader should consult the previous paper \cite{LuMa} for more details and
for a proof of this statement.}. This means that any material body (or
photons) follows along the geodesics (null, in the case of photons) as if the
metric tensor of spacetime was given by the above expression.

{}From the previous calculation we find for the effective geometry the
form

\beq
ds^{2} = g_{00} dt^{2} - g_{11} dr^{2} - r^{2}
( d\theta^{2} + sin^{2}\theta d\varphi^{2} )
\protect\label{b2}
\eeq
with
$$ g_{00} = 1 + \mu.  $$
and
$$ g_{11} = -1 + \nu.  $$

Expanding for $r_{c} << r$ we have

$$ g_{00} \approx 1 -\frac{2 M}{r} - \frac{1}{10}
\left\{ \frac{r_{c}}{r} \right\}^{4} + ...$$
and

$$ g_{11} \approx - 1 -\frac{2 M}{r} -  \frac{M}{r}
\left\{ \frac{r_{c}}{r} \right\}^{4} + ...$$

We see that the modification of the flatness of spacetime, as seen by
matter, beyond the order $O(\frac{M}{r})$,  occurs only at order
$O(\frac{M^{3}}{r^{5}})$. This is radically different from the
results of General Relativity and should provide the basis for a future test.


Before this, however let us look into the
corresponding post-Newtonian parameter of our theory. Using the
isotropic coordinate system (see the Appendix) we can re-write our
effective geometry under the form

$$ ds^{2} = (1 - 2 \alpha \frac{M}{\rho} + 2 \beta \frac{M^{2}}{\rho^{2}}
+ ...)  dt^{2} - (1 + 2 \gamma \frac{M}{\rho}) \left\{ d\rho^{2} +
\rho^{2} ( d\theta^{2} + sin^{2}\theta d\varphi^{2} ) \right\} $$

It then follows from a direct inspection on this form of the effective
geometry the values of the PPN parameters corresponding to our solution. We
can use the associated table in order to compare the values of our model and
those obtained from General Relativity.

\begin{table}
\caption{PPN Parameters}
\label{ext}
\begin{center}
\begin{tabular}{c r r} \hline
          & GR  & LN   \\ \hline
\hline
$\alpha$  & 1   & 1     \\ \hline
$\beta$   & 1   & 1     \\ \hline
$\gamma$  & 1   & 1     \\ \hline
\end{tabular}
\end{center}
\end{table}

A direct inspection on this table led us to conclude that, at the PPN level
of observation, both theories provide the same answer for the efects of the
gravitational field on matter.

\section{Gravitational Energy}
\protect\label{energy}

The fact that we are dealing with a field theory in the conventional
way provide us directly as a by-product, with a well-defined definition of
the energy. We can arrive at the expression of the energy-momentum
tensor either by a direct variation of the associated background
metric or by means of Noether\rq s theorem\footnote{The reader should
note that many differents forms of the energy-momentum tensor of the
gravitational field have been described in the literature. Most of then
suffer from the disease of being pseudo quantities and not true tensor. Some
others, like the one described in \cite{GPP}, are true tensors but have a
hidden gauge symmetry. We note that our expression does not seem to suffer of
any of these difficulties.}.

Using the definition

\beq
{T^{g}}^{\mu\nu} = - \frac{2}{\sqrt {-\gamma}} \frac{\delta L
\sqrt{-\gamma}}{\delta \gamma_{\mu\nu}}
\label{n10}
\eeq
we find, from the above Lagrangian,

\beq
{T^{g}}_{\mu\nu} = - L \gamma_{\mu\nu} + L_{A} \left\{ 4
F_{\mu\alpha\beta}  {F_{\nu}}^{\alpha\beta} + 2
F_{\alpha\beta\mu}  {F^{\alpha\beta}}_{\nu} - 3 F^{\alpha}
F_{\alpha(\mu\nu)} - \frac{5}{2} F_{\mu} F_{\nu} + F^{\epsilon}
F_{\epsilon} \gamma_{\mu\nu} \right\}
\label{n12}
\eeq
Let us specialize this for the case of a static and spherically
symmetric configuration. Since in our theory the Lagrangian is given by
eq. (\ref{Z2}) we have, under the conditions of the above solution,

\beq
L = \frac{b^{2}}{k}
 \left\{ ( 1 - (\frac{r_{c}}{r})^{4} )^{-\frac{1}{2}} -  1 \right\}.
\protect\label{n121}
\eeq

Note that it then follows that $$ L > 0. $$

Finally we obtain, for the density of energy, the expression

\beq
\frac{k}{b^{2}} {T^{g}}_{00} = 1 - \frac{1}{\sqrt{1 -
{(\frac{r_{c}}{r})}^{4}}} +  \frac{32 M^{2} b^{2}}{r^{4}}
 \frac{1}{\sqrt{1 -{(\frac{r_{c}}{r})}^{4}}}
\protect\label{n13}
\eeq
We should point out two important conclusions that follow from this
expression.

\begin{itemize}
 \item{The density of energy of the gravitational field is positive.}
 \item{There exists a true singularity of the field at the point
$r = r_{c}$.}
\end{itemize}

Integrating from $r = R >> r_{c}$, the radius of the compact object,
until infinity a direct simple calculation gives a finite value for the total
energy, provided by

$$ E_{T} = 4 \pi \frac{32 M^{2}}{R}.  $$

This is the value one should expect on Newtonian grounds.

\section{CONCLUSION}
\protect\label{conclusion}
In this paper we continue the exam of a new theory of gravity.
We have found the exact solution of a compact spherically symmetric and
static configuration. We have shown that, although the general system of
equations is highly non linear,
the system reduces, for this symmetry, to a very simple one that allows a
direct integration to be done. An analysis of the properties of our solution
shows many points in common with GR. The PPN parameters
do not allow a distinction between these theories. In order to exhibit a
crucial observable difference between them, we should look for the
propagation of gravitational disturbances. Indeed, (see the Appendix) the
velocity of the gravitational waves is not the same in these two theories.
This should be a clear test to select the good model for gravitational
interaction.

\section{APPENDIX}
\subsection{The Velocity of Gravitational Waves}

In \cite{LuMa} we have shown that the gravitational waves travel in
different cones than the photons. This means that the gravitational waves
feel a geometry of the spacetime that is not the same as the one felt by
matter. The propagation of the gravitational disturbances follow along
trajectories $k^{\mu}$ that satisfy the equation

$$ \left\{ \gamma_{\mu\nu} - \frac{1}{b^{2}} {T^{g}}_{\mu\nu} \right\}
k^{\mu} k^{\nu}  =  0.  $$

This is one of the most important distinctions between the present
field theory of gravity and General Relativity.

\subsection{PPN Formalism}
The standard form of the metric in the PPN form is given by

$$ g_{00} = 1 - 2 \alpha \frac{M}{\rho} + 2 \beta
\left(\frac{M}{\rho}\right)^{2}. $$

$$ g_{ij} = ( - 1 - 2 \gamma \frac{M}{\rho} -
\lambda \frac{M^{2}}{\rho^{2}} ) \delta_{ij}. $$

We make a coordinate transformation to pass from this
isotropic system $(t, \rho, \theta, \varphi)$ to the spherical one
$(t, r, \theta, \varphi)$:

$$ \rho = r \left\{ 1 - \gamma \frac{M}{r} - \frac{1}{2}
(\lambda - {\gamma}^{2} ) \frac{M^{2}}{r^{2}}    \right\}  $$

It then follows that in the new system the metric takes the form
\begin{eqnarray}
{ds}^{2} & = & \left\{ 1 -  2 \alpha \frac{M}{r} -
2 (\beta - \alpha \gamma)
 \frac{M^{2}}{r^{2}}  \right\} {dt}^{2} -
\left\{ 1 + 2 \gamma \frac{M}{r}
+ ( 2 \lambda + {\gamma}^{2} ) \frac{M^{2}}{r^{2}} \right\} {dr}^{2} \nonumber
\\
& - & {r}^{2} ( d{\theta}^{2} + {sin}^{2} \theta d{\varphi}^{2} ). \nonumber
\end{eqnarray}

\subsection{Connections of the Background}

Just for completeness let us enumerate the non-identically null
connections of the background metric \ref{s1} defined by

$$ {\Delta^{\alpha}}_{\mu\nu} \equiv \frac{1}{2} \gamma^{\alpha \beta}
\left\{ \gamma_{\beta \mu,\nu} + \gamma_{\beta \nu,\mu} -
\gamma_{\mu\nu,\beta}  \right\} $$

They are:

$$ {\Delta^{1}}_{22}  = - r. $$

$$ {\Delta^{1}}_{33}  = - r sin^{2} \theta. $$

$$ {\Delta^{2}}_{12}  = \frac{1}{r}. $$

$$ {\Delta^{2}}_{33}  = - sin \theta cos \theta. $$

$$ {\Delta^{3}}_{13}  = \frac{1}{r}. $$

$$ {\Delta^{3}}_{23}  =  cot \theta. $$

\section{CONCLUSION}


\begin{thebibliography}{100}
\bibitem{LuMa} Luciane R. de Freitas and M. Novello in "What is the
velocity of gravitational waves?" preprint, CBPF (1995).
\bibitem{Feynman} R.P.Feynman in Lectures on Gravitation, California
Institute of Technology (1962), unpublished.
\bibitem{Deser} S. Deser, J. Gen. Rel. Grav. 1, 9, (1970)
\bibitem{GPP}  L.P.Grischuck, A.N.Petrov and A.D.Popova, Commun.
Math. Phys. 94, 379 (1984)
\end{thebibliography}
\end{document}